\documentclass{IEEEcsmag}

\usepackage[colorlinks,urlcolor=blue,linkcolor=blue,citecolor=blue]{hyperref}
\expandafter\def\expandafter\UrlBreaks\expandafter{\UrlBreaks\do\/\do\*\do\-\do\~\do\'\do\"\do\-}
\usepackage{upmath,color}
\definecolor{purple}{rgb}{0.5, 0.0, 0.5}
\definecolor{pink}{rgb}{0.87, 0.44, 0.63}

\usepackage{xcolor}
\definecolor{cgaorange}{rgb}{0.98,0.64,0.098}

\definecolor{pipelinedata}{RGB}{68, 114, 196}
\definecolor{pipelineinteraction}{RGB}{0, 176, 80}
\definecolor{pipelinevis}{RGB}{237, 125, 49}
\usepackage{tikz}
\newcommand*\roundedbox[2]{%
\tikz[baseline=(X.base)]%
\node[draw=white, rectangle, semithick, rounded corners=3pt, inner sep=3pt, fill=#2,text=white,align= left](X){#1};%
}

\usepackage{enumerate}

\usepackage{float,tcolorbox}
\newfloat{info@box}{tbp}{loi}%
\makeatother
\floatname{info@box}{Information}
\newenvironment{infobox}[1][]{
   \begin{info@box}%
      \begin{tcolorbox}[colback=white,colframe=cgaorange,title=\textbf{#1},bottomrule=2mm,toptitle=0.5mm,bottomtitle=0.5mm,sharp corners=all]%
}{%
      \end{tcolorbox}%
   \end{info@box}%
}

\usepackage{soul}

\jvol{XX}
\jnum{XX}
\paper{8}

\begin{document}


\title{AI-in-the-loop: The future of biomedical visual analytics applications in the era of AI}
\author{Katja Bühler}
\affil{Vienna Research Center for Visual Computing, VRVis GmbH, Wien, 1220, Austria}

\author{Thomas Höllt}
\affil{Delft University of Technology, Delft, 2600 AA, The Netherlands}

\author{Thomas Schultz}
\affil{University of Bonn, Bonn, 53115, Germany and Lamarr Institute for Machine Learning and Artificial Intelligence}

\author{Pere-Pau Vázquez}
\affil{ViRVIG Group and Universitat Politècnica de Catalunya, Barcelona, 08034, Spain}


\begin{abstract}\looseness-1 AI is the workhorse of modern data analytics and omnipresent across many sectors. Large Language Models and multi-modal foundation models are today capable of generating code, charts, visualizations, etc. How will these massive developments of AI in data analytics shape future data visualizations and visual analytics workflows? What is the potential of AI to reshape methodology and design of future visual analytics applications? What will be our role as visualization researchers in the future? What are opportunities, open challenges and threats in the context of an increasingly powerful AI? This Visualization Viewpoint discusses these questions in the special context of biomedical data analytics as an example of a domain in which critical decisions are taken based on complex and sensitive data, with high requirements on transparency, efficiency, and reliability. We map recent trends and developments in AI on the elements of interactive visualization and visual analytics workflows and highlight the potential of AI to transform biomedical visualization as a research field. Given that agency and responsibility have to remain with human experts, we argue that it is helpful to keep the focus on human-centered workflows, and to use visual analytics as a tool for integrating ``AI-in-the-loop''. This is in contrast to the more traditional term ``human-in-the-loop'', which focuses on incorporating human expertise into AI-based systems.
\end{abstract}

\maketitle

\chapteri{V}isual Analytics has demonstrated its utility in the creation of artifacts that help to solve real-world problems. In the area of public health and clinical medicine, for example, visualization and visual analytics are currently prevalent techniques for knowledge discovery and treatment planning. However, its potential has not been fully harnessed.
AI, and more specifically deep learning (DL) techniques, already play a role in the development of Visual Analytics tools for biomedical data. For example, in the data acquisition and processing stages, DL techniques are crucial to several tasks, such as noise removal, segmentation, etc. But the seamless integration of AI techniques in VA tools has still a long path ahead~\cite{wu2023grand}. Concurrently, AI technology is evolving at an extremely fast pace. The development of transformer-based architectures and along with this, large foundation and language models (LLMs) like chatGPT~\cite{OpenAI_2023} and their evolution towards multi-modal models has revolutionized our understanding of what can be achieved with deep learning. 
More recently, this has evolved into a discussion about the future development of Artificial General Intelligence (AGI). Morris et al.~\cite{Morris2024} categorized recently the current status and potential developments of AGI into "narrow" (i.e. specialist) and "general" AI, and each category into five levels from emerging (comparable to unskilled humans) to superhuman (outperforming all humans). While narrow AI can already achieve superhuman performance to date like AlphaFold3~\cite{Abramson2024}, AGI is still at emerging technology level that cannot yet compete with human performance.
While there is no guarantee that superhuman AGI will be achieved, it is reasonable to anticipate further developments towards lower levels of AGI with high potential to result in transformative shifts in the way we interact with machines, as users, as well as researchers and developers~\cite{Morris2024}.

Taking such developments and trends in AI research as a technological base line, Visual Analytics is facing a radical transformation as a field of research.
While for Visual Analytics, the \textbf{human-in-the-loop} \textit{``has to be the ultimate authority in giving the direction of the analysis along his or her specific task''}~\cite{Keim2007}, in machine learning, the concept of human-in-the-loop commonly refers to the least automated and thus least sophisticated and least desirable solution~\cite{Morris2024}.
We are convinced that AI will become an integrated part of the Visual Analytics loop beyond its core analytics functionality, helping to ultimately design and implement  individualized, adaptive human- and task-centered Visual Analytics solutions for complex data on demand where deep learning-based, and more specifically, generative, analytical and predictive systems collaborate with humans. 

\begin{figure*}
    \centering
    \includegraphics[width=\linewidth]{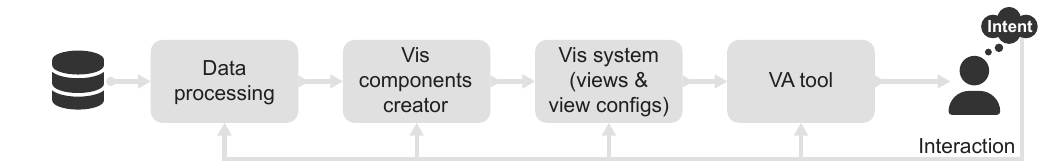}
    \caption{Classical Visual Analytics workflow. To create VA tool, first data is acquired and processed. Then, the visualization components, which are dependent on the task, are created, and used to configure the final Visual Analytics tool.}
    \label{fig:workflow}
\end{figure*}

This Visualization Viewpoint presents our ideas of how deep learning techniques may influence and play a role in the creation and use of Visual Analytics pipelines in the future. We reflect on the current and possible future contributions and implications of incorporating AI into (1) the VA tool creation pipeline and (2) the user interaction with AI empowered VA tools, that is, putting \textbf{AI-in-the-loop} from VA solution development up to an analytical task performed by an end user using these tools. To this end, we rethink the VA tool construction pipeline to add tasks that we envision will be feasible to develop cooperatively with AI technologies, such as LLMs. In addition, we depict our vision of what the role of Assistive AI in context of Visual Analytics for end users should be in the future, and how AI will shape our work as Visual Analytics researchers.

Inspired by previous discussion in context of the Dagstuhl-Seminar 23451
\href{https://www.dagstuhl.de/23451}{"Visualization of Biomedical Data – Shaping the Future and Building Bridges on Biomedical Data"}, our discussion in this Visualization Viewpoint will not primarily focus on standard data analytics cases. Instead, it is driven by real-world, non-trivial requirements defined through data and task complexity as present in the biomedical domain.

\section{Visual Analytics Workflows: present and future}

\subsection{Current VA status}

We focus on the potential effects of AI, and particularly, DL and generative systems in the workflow of the development of Visual Analytics tools. The different steps of such workflows are depicted in Figure~\ref{fig:workflow}. They include the data acquisition and processing, together with the elements that transform such data into visual analytics tools. Overly simplified, these consist of the creation of the different visualization components, the configuration of such components, and the creation of the final Visual Analytics interactive tool.

From those steps, only the first one, data acquisition and processing, has already been established as a stage where Machine Learning (ML) and DL techniques play a crucial role. Tasks such as noise removal, image segmentation, feature extraction, to name a few, are commonly designed with the help of these tools including related methods to enhance transparency and explainability of machine made decisions to end users.
The remaining components of the pipeline are also undergoing alterations under the influence of AI, although these are still largely superficial in nature.\\
Currently, the emergence of LLMs is changing completely the way we develop software. As a result, we envision a future VA tool development pipeline that will be enhanced by generative AI. Those technologies will assist the visualization developer in many of such steps. Not only the design and development will be carried out in collaboration with AI tools, but it will also influence the way how users will interact with these tools, for example, by enabling natural language commands or supporting analytical workflows through intent predictions. 

\begin{figure*}
    \centering
    \includegraphics[width=\linewidth]{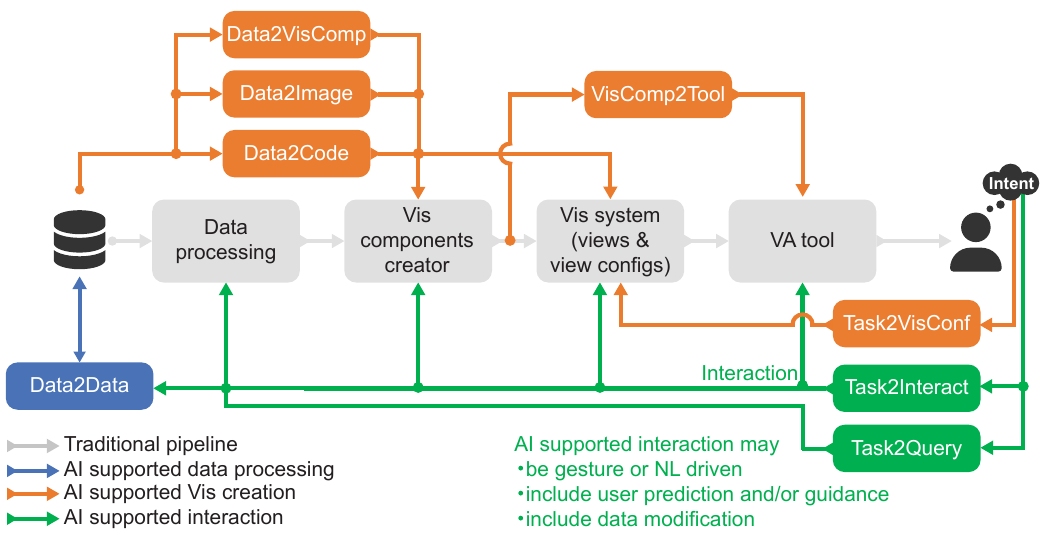}
    \caption{AI-enhanced Visual Analytics workflow construction: In this scheme, we depict different stages of the creation of a VA tool that could be enhanced through the use of novel deep-learning technologies.
    {\color{gray}Gray arrows} correspond to the classical VA pipeline (\autoref{fig:workflow}), {\color{pipelinedata}blue arrows} indicate new AI-supported data processing steps, {\color{pipelinevis}orange arrows} indicate new AI-supported steps in the tool creation pipeline (note: most of the orange arrows are on the top part of the figure, flowing from data to user, but one path on the bottom is based on tasks/user intent and thus flows from the user backwards). Finally, {\color{pipelineinteraction}green arrows} indicate AI-augmented interactions during tool usage and thus replace the classical interaction arrow from \autoref{fig:workflow}.}
    \label{fig:AIworkflow}
\end{figure*}

\subsection{The future of VA workflows}

While the idea of automating the creation of visualizations is not new, most previous systems were constrained to rigid rules, and had limited capabilities. Only recently, with the advent of more advanced natural language technologies and generative models, a greater flexibility has been gained, and this creation process is getting closer to a conversational and intuitive procedure where an AI assistant helps in generating code and solving problems and doubts. On the other hand, there are plenty of examples of language-powered queries to extract information from data. What is new in this area, is the extraction of higher level insights, opposed to concrete queries (e.g., what is the maximum value of a certain variable) directly from the data or the visualization itself. Therefore, tasks such as storytelling, or labelling visualizations are becoming feasible.
Our vision is illustrated in Figure~\ref{fig:AIworkflow}.

We classify the contribution of AI into three different categories, that differ a lot in their nature, but that can be supported in different manners by AI and generative systems, as shown in Figure~\ref{fig:AIworkflow}: (1)~\roundedbox{data}{pipelinedata} acquisition and processing, (2)~components that can aid in the VA \roundedbox{tool creation}{pipelinevis}, and (3)~the components devoted to assisting \roundedbox{interaction}{pipelineinteraction} with such a tool.

The first category includes a set of well-known and established ML methods that are continuously developing to tackle new datasets modalities and increased accuracy. We call those tasks \roundedbox{Data2Data}{pipelinedata}.

Regarding VA tool creation, tasks include: 
\begin{itemize}
    \item \roundedbox{Data2Code}{pipelinevis}: Generation of code for visualization components.
    \item \roundedbox{Data2Image}{pipelinevis}: Extraction of visual depictions from the data. 
    \item \roundedbox{Data2Viscomponent}{pipelinevis}: Besides individual charts or images, other elements such as filters, menus, etc., need to be defined to create a VA tool. These basic components could also be created by generative models, since they do not differ much from other code required to build charts.
    \item \roundedbox{Task2VisConfig}{pipelinevis}: All the components in a VA tool require configuration. These include changing size, colors, palettes, etc. 
    \item \roundedbox{Viscomponent2Tool}{pipelinevis}: The creation of a VA tool involves linking many visualization components, and laying them out in a dashboard. 
\end{itemize}
All tool creation tasks have in common that they are not mere transformations of input data. On the contrary, they require generative capabilities. 
Even though some applications such as Lida~\cite{Dibia_2023} can already generate pieces of code to visualize simple datasets, they are restricted to very basic charts. Thus, they do not generalize to the complex input medical or biological pipelines require. As LLMs evolve and specialize, direct generation of custom visualization code adapted to the current dataset will be feasible. These code snippets rely on the use of visualization libraries to handle the final visualization. 

\roundedbox{Data2Image}{pipelinevis} is a more sophisticated version of \roundedbox{Data2Code}{pipelinevis}. In this case, the AI component will be able to transform data directly into an image. When transforming data into visualization code, no care is taken on the final aspect of the visualization. The configuration of the visual variables is commonly left to the library itself, which often has default parameters that work for a wide range of situations. Selection and fine-tuning of the aesthetics of the final visualization is a task that is currently only done by humans, but could be taken over by large complex systems if enough data (and maybe guidelines) to train them were available.

Each visualization component needs to be adapted to the VA tool: from modifying its size to changing palettes, many aspects can be fine-tuned. Currently, this is carried out by the designers, but, like in previous cases, in the future they may end up being tasks that can be triggered by the designers, and performed by AI-supported tools.

Like in the previous case, we believe that the process of creating a tool by mixing and linking different visualization components (\roundedbox{Viscomponent2Tool}{pipelinevis}) will be achievable in the near future with the support of generative systems. 

Concerning interaction, we anticipate that the pipeline will benefit from the advances in Natural Language Processing, notably LLMs. Furthermore, they will also integrate other DL techniques to include both more \emph{classical} tasks, such as querying directly the data, and more high-level procedures such as AI guidance for the interaction with the visualizations using natural language. These tasks consist of:
    
\begin{itemize}
    \item \roundedbox{Task2Query}{pipelineinteraction}: Execution of queries about data (e.g., find values) or visualization insights (e.g., determine insights, correlations, etc.) using natural language. Here, AI techniques will improve the understanding of complex queries or user intentions. But they will also be able to propose or modify queries to help users interpret the data, since they will be able to obtain a more profound understanding of the input dataset. 
    \item \roundedbox{Task2Interact}{pipelineinteraction}: Interaction with the tool using natural language, such as for filtering, selection, or getting values. Again, AI can help either by facilitating the interaction (e.g., predicting the users' movements) or by hinting at potential tasks of interest. 
    
\end{itemize}

Natural language has already been demonstrated for simple interactions with data, such as finding values. More elaborate interactions are feasible with sophisticated language models that can translate questions into the creation of filters, aggregations, or data extraction tasks. 
However, more complex data, with multiple dimensions, or where queries require understanding of high-level structures (e.g., anatomy) or features (e.g., a visual cue), have still not been showcased in the literature. In the future, this task, \roundedbox{Task2Query}{pipelineinteraction}, can be further assisted by generative models if these improve their analysis capabilities and can point to relevant features or regions of interest. This might eventually include storytelling, analysis of temporal data evolution, extraction of relevant insights for a certain task, etc.

\roundedbox{Task2Interact}{pipelineinteraction} is concerned with direct manipulation of complex datasets. This can be time-consuming and cumbersome for complex scenarios. AI might guide the interaction both by identifying regions of interest, or by creating filters that involve multiple variables and thresholds, for example. But also anticipating users' intentions, or completing their actions. This would reduce the time necessary for data exploration, as well as guarantee that relevant facts are detected.

\begin{figure*}
    \centering
    \includegraphics[width=\linewidth]{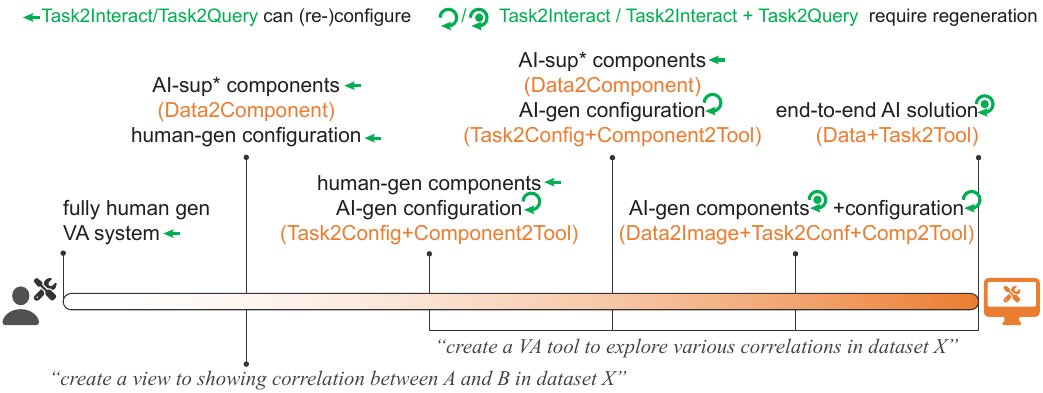}
    \caption{Spectrum of AI in the VA tool generation process. AI-sup* indicates AI supported tasks, such as creating code fragments, that also means that the output is a configurable component that can be directly edited, configured, and interacted with, by a human user. AI-gen(erated) indicates a closed AI solution, that cannot be modified. Note: positions indicate only the order or increasing automation but do not quantify their differences. For example, AI-gen components+configuration is very close to end-to-end, with the difference that it is created step-wise and parts can be modified which is not possible for an end-to-end solution.}
    \label{fig:AIforVAtoolgenSpectrum}
\end{figure*}

The different parts of the pipeline can be mixed and matched to create various levels of automation (see Figure~\ref{fig:AIforVAtoolgenSpectrum}). On the one end we have a fully human-generated and operated system as is common today. Components are manually implemented, combined and configured for the task at hand. On the other end of the spectrum a complete VA system could be generated as an end-to-end AI solution, based on a prompt, specifying data and intent only. In between are combinations of human- and AI-generated components and configurations. As discussed above, we already see early versions of AI-supported components (e.g., creating code for a view, using an LLM) configured by humans, or AI-generated configurations (e.g., through interaction with natural language). However, generating deeply integrated tools with a high level of flexibility is still far off. Of note, interaction can be added manually or AI-supported in any configuration, however, the closer we get to end-to-end solutions, the more regeneration might be required, posing challenges for continuity. Consider an intent-driven interaction for a volume view: \textit{``Make soft-tissue transparent.''} If the transfer function is created as a configurable component of the view, the human operator or AI can directly manipulate the transfer function based on such a prompt and the resulting visualization will adapt. In an end-to-end volume view the transfer function might be a purely abstract concept and changing it requires a complete regeneration of the visualization.

Ensuring the integrity, explainability and transparency of AI-generated VA workflows, including all their components, is crucial, but will remain a major challenge in the coming years, both from an AI methodology and from an AI-user communication point of view. Even the most advanced LLMs still hallucinate and fail to provide trustworthy and faithful explanations and reasoning for generated solutions to complex tasks (see \cite{bubeck2023} and the controversies surrounding this paper). It is currently an open question how end users or developers can efficiently ensure, that the created tools will work correctly emphasizing the ongoing importance of XAI research. Rather than waiting for the remote hope that some future AGI might be able to reason correctly about every decision it makes, we believe that the visual analytics community should realize its immediate potential to making existing systems more explainable.

\section{Challenges in Biomedical Data Analytics}

Independent from any Visual Analytics applications, deep learning is already an established method in biomedical informatics. Applications range from small sub-tasks like image registration, data filtering or data classification tasks up to powerful models providing for example medical doctors with end-to-end diagnostics or life scientists with fully automated pipelines for protein discovery~\cite{Abramson2024}. But even if the visions of a super-human generalist AI like described by Morris et al.~\cite{Morris2024} or generalist medical AI by Moor et al.~\cite{Moor2023} seem to indicate at first sight that interactive data exploration and analytics with Visual Analytics is becoming obsolete and data-driven decision-making might be fully automatized in its extreme case, this can be questioned due to several reasons. 
\textbf{(1) Data in the biomedical field are highly dynamic as data acquisition is, especially in the life sciences, undergoing constant technical evolution and innovation. } Novel imaging technologies are for instance developed not only in terms of increased scales, imaging quality and speed, but also e.g., in terms of novel markers highlighting specific biological function at the molecular level or technology capturing transcriptomics and metabolomics in spatial samples. This dynamic is making data analytics very demanding, as completely novel types of before potentially impossible observations exhibiting novel features would have to be handled reliably by an AGI (or lower-level existing AI solutions). 
\textbf{(2) Data in the biomedical field often lacks standardization, missing or uncertain data is common.} Thus, automated processing based on original raw data becomes challenging, requiring expert and potentially also insider knowledge not previously exposed to any AGI. 
\textbf{(3) Data complexity and dimensionality is constantly increasing and together with it the complexity of questions to the data and related analytical tasks.} Beyond standardized diagnostic tasks, this is a highly dynamic field driven by the dynamics in data acquisition technologies (see (1)) and
\textbf{(4) Entirely new situations may arise where we are confronted with unprecedented data, such as during the Covid 19 pandemic.}

Coping with these challenges autonomously requires a broad set of highly specialized ``emerging''\footnote{Emergent properties of an AI are capabilities not explicitly anticipated by a developer} skills that will hardly be reached by any AGI in the near future. Furthermore, by definition the idea of a generalist AI pushes humans into a more passive role in which the AI delivers the results, which the human finally only approves or rejects. Such an end-to-end scenario could easily fail in complex data environments as described above, in which the learnt skills of an AGI are not necessarily sufficient to arrive at reliable results. It also contradicts the concept of human centered AI being a central element of current ethical and regulatory frameworks (see section below). 

The combination of AI-driven data analysis with visual analytics as an interface between the user and the AI establishes a different concept of cooperation between humans and AI. It comes with the opportunity to create an environment that can react more flexibly to dynamic, complex data environments and tasks and supports users in drawing their data-based conclusions as discussed above, by bringing the AI-in-the-loop. 
Therefore, we envision the emergence of AI companions that can create end-to-end VA tools based on simple prompts. However, just like for the use of AI, the biomedical domain poses significant challenges for the the development of such AI-in-the-loop VA tools. Extending on the challenges laid out above, we observe that \textbf{(5)
VA researchers frequently encounter a scarcity of biomedical training data, which presents a significant challenge in developing unbiased and stable AI systems.}
The high variability, complexity and volatility of data (see (1), (2), and (3)), as well as complete lack of data e.g. in case of rare or new diseases (4) requires the modification of  both, data processing algorithms and the inclusion of new analyses and interpretations into an VA system. However, data covering this knowledge is not or only sparsely available to (re-)train any assistive systems.
\textbf{(6) There is only a limited availability of VA code for training or tuning AI models.} The majority of VA tools are proprietary, and their code private. Therefore, it cannot be used to train new AI systems. This limits the potential of generative tools to build specific biomedical VA applications.
\textbf{(7) Complex interactions with complex biological or medical models in exploratory analysis tools are difficult to mimic.} Biomedical experts are highly specialized, and the interactions are data-specific and highly variable between experts. As such, user behavior is difficult to train and predict and generated interactions are hard to \textbf{validate}. This becomes even more challenging when designing for \textbf{inclusivity} to support heterogeneous user groups, from bioinformaticians and medical experts to non-specialized medical personnel.
\textbf{(8) In compliance with the diverse legal frameworks on the use of AI, high-risk systems must be transparent and should not replace human sovereignty over a decision.} Moreover, these decisions should be explainable. This needs to be considered through the whole VA system design ensuring that every AI-generated subtask that is substituting human actions must be law-compliant, reliable, and the underlying reasoning can be monitored and validated.

\section{Dangers, Ethical, and Legal Considerations}

The tight integration of AI into visual data analysis pipelines and applications for biomedical purposes in professional environments requires careful consideration of ethical and safety issues. Moreover, the use of AI technologies is not free from dangers \cite{bender2021dangers}. Errors, mistakes, or inaccuracies in the development and use of AI could have serious implications for the quality and reliability of results, with potential legal and/or financial consequences, e.g., if a patient is misclassified for a disease or if inaccurate or incorrect predictions slow down the process of discovering new drugs.  However, the black box nature and unreliability of powerful LLMs and multi-modal foundation models trained on massive sets of data, often lack the necessary detail in model and data documentation \cite{bender2021dangers}, making it difficult to ensure the implementation of high ethical and safety standards in downstream applications. 

Recently, several policy papers and resolutions have been published worldwide to provide a regulatory and ethical framework for the use of AI such as the \href{https://www.unesco.org/en/articles/recommendation-ethics-artificial-intelligence}{UNESCO Ethics of AI}, the \href{https://oecd.ai/en/ai-principles} {OECD Principles for trustworthy AI} which are partially reflected in national regulatory frameworks like the \href{https://ai.gov, https://www.whitehouse.gov/briefing-room/presidential-actions/2023/10/30/executive-order-on-the-safe-secure-and-trustworthy-development-and-use-of-artificial-intelligence/}{US Executive Order on the Safe, Secure, and Trustworthy Development and Use of Artificial Intelligence}, the Chinese \href{https://digichina.stanford.edu/work/translation-internet-information-service-algorithmic-recommendation-management-provisions-effective-march-1-2022/}{Internet Information Service Algorithmic Recommendation Management Provisions}, or the  \href{https://digital-strategy.ec.europa.eu/en/policies/regulatory-framework-ai}{EU AI Act}. A worldwide overview on AI legislation which is regularly updated is provided e.g. by \href{https://iapp.org/resources/article/global-ai-legislation-tracker/}{The International Association of Privacy Professionals (IAPP)}. 

One of the key statements in most declarations is the need for human oversight and fairness. 
While the use of AI in research is often not as strictly regulated, visualization researchers should be aware of these frameworks when designing AI-supported applications. Considering reliability, legal and ethical issues throughout the project has the potential to increase adoption by domain experts and/or industry and to enhance the impact of the research results. One framework is provided by the {\it Ethics by Design} approach, which covers the design, development, deployment, and use of AI-based solutions. {Ethics by Design} considers the need for human agency; privacy, personal data protection and data governance; fairness; individual, social and environmental well-being; transparency; accountability and oversight.

 Research thrives on pushing the boundaries, but not all what is technically possible might be finally accessible and usable. The knowledge on  potential regulations of foundation models is a prerequisite to create sustainable and usable results. For instance Metas' llama3 model is currently (June 2024) not available in Europe due to regulatory reasons \footnote{https://about.fb.com/news/2024/06/building-ai-technology-for-europeans-in-a-transparent-and-responsible-way/}.

 \section{The future role of visual analytics in biomedical applications}
 In the following discussion on the future of VA in biomedical applications, we focus on scenarios with complex data and analytical tasks. Therefore, we skip simple cases related to narrow, supportive AI for accelerating non-critical tasks, like optimized data loading, filtering, etc. 

The quality of the result of a decision making process is the ultimate criterion for a performant human-AI teaming, in particular in critical environments like medicine or life science research. Here, very much following Keim et al.~\cite{Keim2007}, agency has to fully remain with a human expert, who takes responsibility for diagnostic and treatment decision-making in the medical domain, or for the scientific integrity of the overall reasoning and conclusions in a research context. Consequently, the human must have an opportunity to audit all algorithmic results and decisions. During this process Visual Analytics remains to play an inevitable role: It provides tools for comparative analysis (e.g., to explore the AI-based classification results of a cancerous patient in context of multi-modal data of a related patient cohort), for understanding why an AI-based assistant or second reader made a specific suggestion (e.g. by translating potential explanations of an AI into a more human interpretable form by providing a visual representation of context), or for assessing uncertainty in algorithmic results, in particular, when real-world datasets differ from those that the system was trained on (by visualizing domain shift or uncertainty of decisions). 

``Generalist'' medical AI systems~\cite{Moor2023}, combine the ability to integrate multi-modal data (including text, images, clinical, and -omics) with broad medical knowledge. These systems should support complicated general analytic tasks, such as generating full radiological reports, or providing real-time surgical or bedside decision support, rather than improving or solving some narrowly defined sub-task. Even though training foundation models on large and diverse datasets has significantly improved their ability to generalize across tasks and data characteristics compared to previous AI models, they are currently far from reliable enough to be deployed autonomously within critical applications. Fusing the concept of ``Collective Intelligence'' (CI) with AI can be one way to overcome these limitations~\cite{cui2024ai}. CI refers to the joint problem-solving performance of several individuals, more recently also in combination with analytical capabilities of machines, and is a trend in current interdisciplinary AI research.

In all these scenarios, Visual Analytics will remain an ideal tool for shaping human centered interaction with the AI and other experts. If complex data is underlying a question to “the” A(G)I, it still has to be defined and potentially selected from different large data sets or collections, requiring potentially also quality control of the data defining additional steps. Doing this only with a textual prompt might be by far less efficient than having a visual tool, allowing a much more intuitive definition of a respective task. Exploration of the unknown for hypothesis building based on highly complex data is another important use case requiring, e.g., comparative visualizations and interactive data exploration and iterative query refinement. Effective, intuitive, and trustworthy communication of results and joint analysis is an important prerequisite to realize collective human-machine intelligence. 

Enabling \textbf{AI-in-the-loop} solutions that effectively couple the power of AI driven data analytics with Visual Analytics to enhance human abilities to cope with the challenges in biomedical data analytics is a highly promising yet demanding endeavour, that will remain an active field of research to solve major challenges including question how to cope with data and task complexity and their dynamics across all levels of the application affecting both, technical aspects as well as user interface and interaction design; or how to shape human-AI teaming as the dense integration of AI-driven data analytics with Visual Analytics for such complex data and tasks is non-trivial. The question of how a user communicates and interacts with an AI and how the AI is communicating with the user in a transparent and explainable manner tailored to the user and her task is an open question, that is additionally complicated in complex scenarios. This does not only hold for VA research, but is also an open question addressed by AI research. 

\section{Conclusion -or- Do robots dream of AI-generated sheep?}
We anticipate that the specific challenges of biomedical applications, which include heterogeneity and limited availability of data and code examples for training, the complexity and specialized nature of the required domain knowledge, the complexity and dynamics of analytical tasks, the permanent proliferation of new acquisition devices and data types, combined with high standards for reliability and trustworthiness, is still slowing down the practical adaptation of AI-based solutions compared to other domains with lower stakes, such as generating content for recreational purposes. However, we believe that, despite these challenges, AI will ultimately lead to profound transformations also in the biomedical domain, and we advocate shaping these changes according to an “AI-in-the-loop” paradigm, in which visualization plays an important role. In contrast to the frequently used term “human-in-the-loop”, which considers humans as a source of information within an AI-based system, the term “AI-in-the-loop” emphasizes that the ultimate focus, responsibility, and control should remain with human experts who use AI-based systems to make faster or better informed clinical decisions, or as a powerful new tool for scientific research.

According to recent estimates, even though the Food and Drug Administration has approved more than 200 commercial radiology AI products, the corresponding U.S. market penetration is still only around 2\%~\cite{Rajpurkar2023}. We expect that our proposed “AI-in-the-loop” approach, which is centered on supporting specific biomedical workflows by a suitable integration of visualization and AI, will more effectively close this gap than “human-in-the-loop” approaches, which are centered on solving specific AI tasks by accounting for human feedback. While AI assistants should make implementation and prototyping of future visual analytics systems substantially easier and more efficient, implementing the “AI-in-the-loop” approach still involves classical stages of visualization design, such as data and task abstraction, that are unlikely to be fully automated in the foreseeable future. Consequently, we expect that, as long as it actively tracks progress in artificial intelligence, and makes use of the opportunities it presents to domain experts as well as tool builders in the spirit of the proposed “AI-in-the-loop” approach, biomedical visual analytics will continue to play a central role in the future.

The inclusion of AI assistants in the development of VA tools will fundamentally change how we work as researchers. Code development implementing traditional, well-known procedures (reading models, creating UIs$\ldots$) will be increasingly taken over by assistive systems. But this only works for tasks that have been part of the training set of the generative models. Researchers will still need to develop new algorithms that deal with new tasks or new modalities of data, design interfaces to deal with new kinds of information or implement new tasks, creation of new tools for overseeing the performance of AI models, development of new tools for monitoring and validating the new VA software. However, the availability of public, labelled datasets of VA tools (or procedures) will be scarce due to multiple reasons, such as the proprietary nature of most of the software, its complexity, etc. Therefore, creation of AI assistive software for Biomedical VA tools will be much slower as compared to other domains, such as generative systems for image creation. 

The pipeline we have proposed in Figure~\ref{fig:AIworkflow} is plausible, and likely will be part of our work environments in a few years, as we move further to more and more capable generative AIs (Figure~\ref{fig:AIforVAtoolgenSpectrum}). However, the possibilities of generative tools go beyond that. And the prospect of AI companions that help us in our work is very feasible. In the future, these will be pervasive, and the development of VA tools will be fundamentally changed by them. 

The rise of AI will not threaten our work as Visual Analytics researchers but provides a unique opportunity for boosting innovation with many open research questions (see box ``Research Opportunities'').

\begin{infobox}[Research Opportunities]

\textbf{Visualization Foundations}
\begin{enumerate}[\indent {}]
\itemsep0.4em
\item Analysis of cognitive and psychological needs and consequences of AI-in-the loop both on human users and the workflows.
\item Enhancing and leveraging the “expertise” of multi-modal AI models towards an artificial visualization expert.
\item Integration of Visualization knowledge into AI models. 
\item Broadening the theory on coupling human creativity and expert knowledge with AI capabilities. 
\end{enumerate}

\vspace{2mm}
\textbf{AI for VA Tool Development}
\begin{enumerate}[\indent {}]
\itemsep0.4em
\item Low to no-code VA tool development, with a focus on security, reliability, transparency, and ethics.
\item Integration of different interaction modalities with AI to build and support VA tools.

\item Creation of VA tools from scarce data or code examples (few-shot VA tool generation).
\end{enumerate}

\vspace{2mm}
\textbf{AI Integration in the VA Workflow}
\begin{enumerate}[\indent {}]
\itemsep0.4em
\item AI-enhanced workflows, adapting to different user profiles (skills, knowledge) and previous and current interaction with the system.
\item Shifting towards multi-modal interaction with the AI (and the data) to jointly complete a data-driven task.
\item Designing new AI-in-the-loop interfaces at all levels.
\item Visual AI-to-human communication methods that provide sufficient explanations to support accurate and reliable  decision making.
\item Quality assessment of AI-in-the loop.
\end{enumerate}

\end{infobox}

\section{ACKNOWLEDGMENTS}
This work has benefited from Dagstuhl Seminar 23451 \href{https://www.dagstuhl.de/23451}{"Visualization of Biomedical Data – Shaping the Future and Building Bridges on Biomedical Data"}.
Katja B\"{u}hler is supported by VRVis which is funded by BMK, BMAW, Styria, SFG, Tyrol and Vienna Business Agency in the scope of COMET - Competence Centers for Excellent Technologies (879730) managed by FFG. Thomas H\"{o}llt is supported by the Sector Plan B\`{e}ta en Techniek at TU Delft. Pere-Pau Vázquez has been supported by PID2021-122136OB-C21 from the Ministerio de Ciencia e Innovación, Spain, by 839 FEDER (EU) funds, and 2021 SGR 01035 by Generalitat de Catalunya.

\def\refname{REFERENCES}
\bibliography{paper}

\begin{IEEEbiography}{Katja B\"{u}hler}{\,} is the Scientific Director of the Vienna Research Center for Visual Computing (VRVis GmbH) in Vienna, Austria, where she also serves as Head of the Biomedical Image Informatics Group. Furthermore, she is board member of the Association for the Promotion of Digital Humanism, member of the management board of Austrian Bioimaging. Her research interests lie in the combination of Visual Computing and AI with the aim of facilitating human-centred  access to data, particularly in the context of supporting complex image-based decision-making processes within the biomedical domain. B\"{u}hler holds a doctoral degree ($Dr.^{in}techn.$) in computer science from TU Wien, Austria and has published over 100 peer-reviewed papers. She is a member of the Eurographics Association (EG), the German Association for Computer Sciences (GI), the Confederation of Laboratories for Artificial Intelligence Research in Europe - (CLAIRE), the Austrian Society for Artificial Intelligence (ASAI), and the Austrian Neuroscience Association (ANA). Contact her at buehler@vrvis.at.\vadjust{\vfill}
\end{IEEEbiography}

\begin{IEEEbiography}{Thomas H\"{o}llt}{\,} is an Assistant Professor in the Computer Graphics and Visualization group at TU Delft, Delft, The Netherlands. His research interests are in Visualization and Visual Analytics, with a focus on bio-/medical applications. He received his PhD from the King Abdullah University of Science and Technology, Thuwal, Saudia Arabia in 2013. Dr. H\"{o}llt published over 50 peer reviewed publications including the winning entry for the Dirk Bartz Prize for visual computing in medicine in 2019. He is a member of the EUROGRAPHICS association which he also serves as publicity and online chair. Contact him at t.hollt-1@tudelft.nl.\vadjust{\vfill}
\end{IEEEbiography}

\begin{IEEEbiography}{Thomas Schultz}{\,}is a professor at the b-it and Department of Computer Science at the University of Bonn, Germany, a principal investigator at the Lamarr Institute for Machine Learning and Artificial Intelligence, and head of a research group on Visualization and Medical Image Analysis. His research integrates quantitative image analysis, machine learning, and interactive visualization, with a current focus on applications in neuroimaging and ophthalmology. Dr. Schultz received his doctoral degree from Saarland University, Germany, after working on a medical visualization topic at the MPI for Informatics. He is a member of the German Association for Computer Science (GI), and its divisions on Visual Computing in Biology and Medicine, and Computer Graphics. Contact him at schultz@cs.uni-bonn.de.\vadjust{\vfill}
\end{IEEEbiography}

\begin{IEEEbiography}{Pere-Pau V\'{a}zquez}{\,} received the graduate degree in computer science and the PhD degree in software from the Universitat Politecnica de Catalunya,
in 1999 and 2003. He is currently an associate professor with the Computer Science Department, Universitat Politecnica de Catalunya in Barcelona.
He is member of the Research Center for Visualization, Virtual Reality, and Graphics Interaction (ViRVIG) and the Center for Research in Biomedical Engineering in Barcelona (CREB). His current interests are mostly related to
visualization of large data sets, the intersection of AI and visualization, perception, and
interaction in virtual reality environments. He has published over 100 peer-reviewed papers. He is member of the EUROGRAPHICS association where he currently serves as secretary assistant. He is also member of the Eurovis Steering Committee. Contact him at pere.pau.vazquez@upc.edu.\vadjust{\vfill\pagebreak}
\end{IEEEbiography}

\end{document}